# INFLUENCE OF THE FEEDBACK FILTER ON THE RESPONSE OF THE PULSED DIGITAL OSCILLATOR

*M. Domínguez[1], J. Pons[1], J. Ricart[1], J. Juillard[2], E. Colinet[2]*

[1]MNT, Universitat Politècnica de Catalunya. Barcelona, Spain.
[2]Département Traitement du Signal & Systèmes Électroniques, SUPELEC. France.

**ABSTRACT**

This paper introduces a new feedback topology for the Pulsed Digital Oscillator (PDO) and compares it to the classical topology. The 'classic' or single feedback topology, introduced in previous works, shows a strong behavior dependence on the damping losses in the MEMS resonator. A new double feedback topology is introduced here in order to help solving this problem. Comparative discrete-time simulations and preliminary experimental measurements have been carried out for both topologies, showing how the new double feedback topology may increase PDO performance for some frequency ranges.

## 1. INTRODUCTION

The use of resonant MEMS devices working near their resonance is widely extended today to a growing number of applications, which include RF components, chemical and gas sensors, accelerometers, gyroscopes, actuators, etc. MEMS resonators are an attractive option to fix the oscillation frequency in such applications because of their high quality and frequency stability propperties. However, most actuation schemes for MEMS (electrostatic, thermoelectric, etc.) are non-linear and therefore the design of large-signal oscillators is not simple. In this way, several oscillator topologies have been proposed in the past; most of them linearize the response of the MEMS by applying a suitable bias voltage and therefore the displacement of the resonator is in the small signal range. On the other hand, large-signal oscillators proposed in the literature exhibited chaotic behavior [1].

The Pulsed Digital Oscillator (PDO) [2-3] is a set of sigma-delta based topologies for MEMS that was originally proposed by the authors to overcome these difficulties so that non-chaotic oscillation waveforms can be obtained even working in the large signal range.

This work introduces and evaluates a new feedback loop structure for the PDO, called *double feedback*. The goals of this new architecture are to obtain better results when damping losses in the MEMS resonator are relevant and to enhance the practical oscillation range. To this effect, the main features of the classical, or *single feedback*, PDO architecture are reviewed in section 2, whereas section 3 introduces the *double feedback* structure and compares it to the classical one through simulations. Finally, section 4 extends this comparison with experimental results.

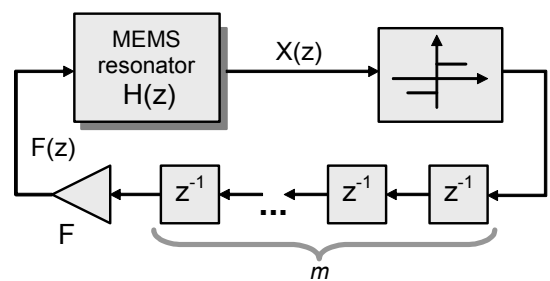

*Figure 1. Block diagram of the "classic" m-delay pulsed digital oscillator (PDO).*

## 2. PDO FUNDAMENTALS

### 2.1. Architecture and key features

The classical PDO architecture is depicted in Figure 1. We can see that it has a simple and relatively easy to implement structure which includes the MEMS device plus a feedback actuation loop composed by a sign detector, a variable number of delay blocks and an amplifier. Some key features of the PDO are:

a) Very short pulses (i.e. deltas) of force of constant amplitude F are supplied to the resonator. This actuation scheme avoids the above mentioned non-linearities present in MEMS actuation. This actuation scheme has been analyzed, extended to other MEMS-based applications and its performance has been evaluated for an accelerometer system in a recent work [4].

b) The PDO is a sampled circuit and, assuming a typical two-parallel plate MEMS resonator structure, at each sampling time it is only checked if the moveable plate is above or below its 'zero', or rest, position. Such simple position measurement requirements enormously simplify the design and implementation of the MEMS device and of the circuit needed to manage the feedback signal.

c) The PDO is a digital oscillator and, in over-sampling conditions, the oscillation frequency ($f_{OSC}$) is directly extracted from the spectrum of the bit stream at the output of the sign detector.

d) The PDO exhibits 'perfect' sinusoidal oscillations at the natural frequency of the resonator ($f_0$) for certain sampling frequency ($f_S$) sets [3]. It has also been demonstrated that the energy transfer from the electrical domain to the mechanical one is maximized in such 'perfect' sample frequencies.

e) When the damping losses in the MEMS resonator are relevant, and due to the presence of the 1-bit quantizer in the feedback loop, the oscillation frequency ($f_{OSC}$) as a function of the sampling ratio ($f_0/f_S$) exhibits a 'fractalized' shape [2,3], which looks rather similar to a distorted version of the well-known devil's staircase fractal [5]. This effect has also been recently reported for a very similar circuit topology in [4].

f) We have recently demonstrated that the PDO can also work very well with sample frequencies below the Nyquist limit [6]. This extends the application scope of this kind of oscillator to wider frequency ranges. However, the effects of the damping losses mentioned above become important when working in deep under-sampling conditions.

**2.2. Basic PDO theory**

Let us now focus in the PDO structure shown in Figure 1 (for m=1), hereafter called *single feedback*. By assuming a typical 1D mass-spring-damping model for the parallel-plate MEMS resonator [7] and applying a linear analysis, it can be easily obtained that the structure produces the following normalized digital oscillation frequency [2,3],

$$f_D = \frac{1}{2\pi}\cos^{-1}(e^{-\rho 2\pi f_0/f_S}\cos(2\pi\frac{f_0}{f_S}\sqrt{1-\rho^2})) \quad (1)$$

being the digital oscillation frequency $f_{OSC}=f_D f_S$, and $f_S$ the sampling frequency, $\rho$ the dimensionless damping factor and $f_0$ the natural frequency of the MEMS resonator. These parameters are related to the well-known ones of the MEMS model according to,

$$\rho = \frac{b}{2\sqrt{km}} \qquad \omega_0 = 2\pi f_0 = \sqrt{\frac{k}{m}} \quad (2)$$

where *m* is the mass of the moveable plate, *k* is the spring factor, or stiffness, and *b* is the damping factor.

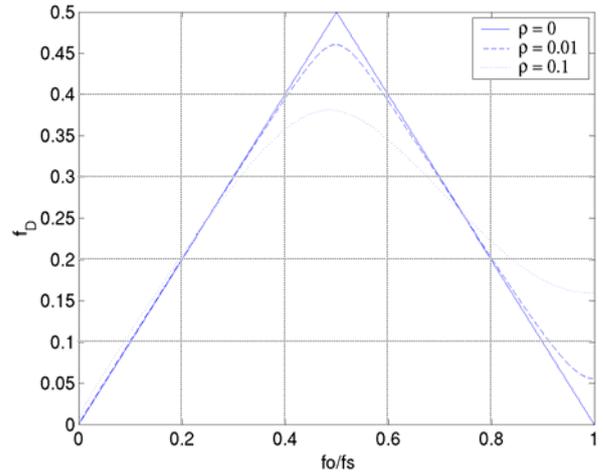

*Figure 2. Results of the linear analysis of the single feedback structure for three different values of the damping losses.*

Figure 2 shows the relationship between $f_D$ and the sample ratio $f_0/f_S$ for three different values of the damping factor after equation (1) for a given MEMS device parameter set [8] and three different values of the damping losses. It can be clearly seen that when damping losses are small, the oscillation frequency closely follows the evolution of the natural frequency of the resonator. In presence of heavy losses, though, the actual response of the oscillator becomes poorer and $f_D$ can clearly depart from the natural frequency of the resonator.

Extensive discrete-time Matlab simulations that have been carried out for the same reference cases reveal that, in the presence of heavy damping losses, the PDO response becomes fractalized and that $f_D$ clearly departs from the natural frequency of the resonator, in particular for sampling frequencies close to the Nyquist limit (see Figure 3). This last effect is not only due to the presence of the fractal, but also because of the tendency marked by equation (1).

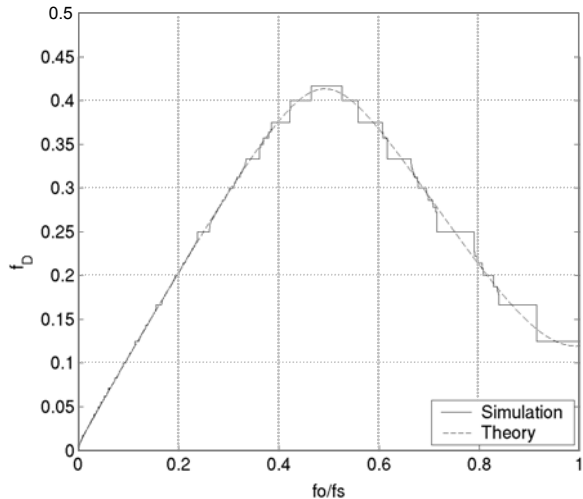

*Figure 3. Discrete-time simulation results versus linear analysis results for a single feedback PDO topology and relevant damping losses ($\rho=0.05$).*

## 3. DOUBLE FEEDBACK TOPOLOGY

In order to obtain a reduction of the influence of the damping losses on the PDO performance, a review of the linear analysis which led to equation (1) has been carried out. From this point of view it can be concluded that such reduction could be obtained by adding some digital filtering in the feedback loop.

According to this and in order to preserve the overall simplicity of the architecture and its design constraints, a new oscillator topology, called *double feedback*, is proposed here. It includes two feedback paths of different delays (see Figure 4). This double feedback structure works somehow like a low order FIR filter.

Let us also notice that with this structure the force pulses applied to the MEMS can take up to three different values {+2F, +F, 0}, instead of the two possible values {+F, 0} which are possible with the single feedback architecture.

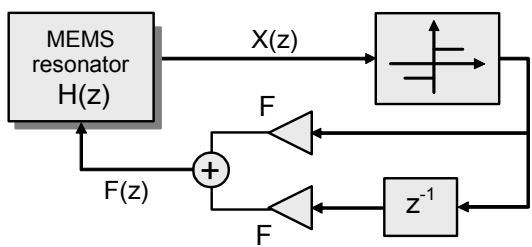

*Figure 4. Topology of the double feedback PDO.*

Discrete-time simulation results obtained for the same reference cases as in Figure 3 but for a double feedback PDO are shown in Figure 5. After a comparison of the results of Figures 3 and 5, it can be easily concluded that the double feedback approach produces a wider range of oscillation frequencies which are closer to the natural frequency of the resonator than in the single feedback case, even in presence of heavy damping losses. In particular, the sample ratio values around the Nyquist limit ($f_0/f_S=1/2$) produce results similar to the 'no-losses' ($\rho=0$) case However, the fractalized behaviour still remains, but becomes more evident for the deep over-sampling and deep undersampling ranges.

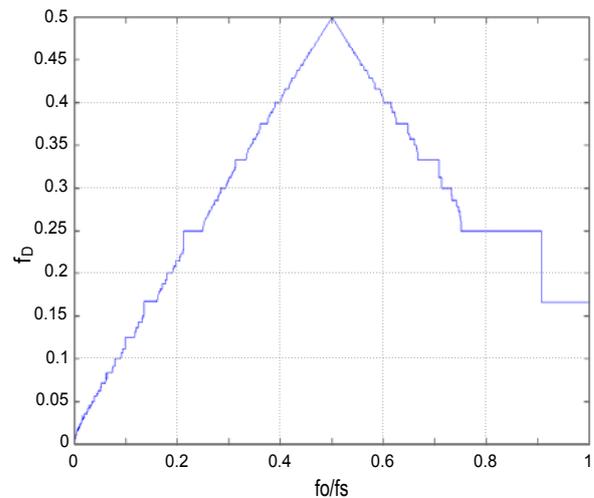

*Figure 5. Discrete-time simulation results for a double feedback structure PDO topology and relevant damping losses ($\rho=0.05$).*

## 4. EXPERIMENTAL RESULTS

The promising perspectives provided by the double feedback architecture about a reduction of the effects of the damping losses on the oscillator performance for some frequency ranges have been checked in a first set of experimental measurements.

### 4.1. Measurements setup

The measurements setup is basically the same one that was used in recent works [2,6] but with the changes in the digital section needed to implement and select between a single and a double feedback loop.

The MEMS resonator comes from the same batch previously described in [2] and it is a cantilever with thermoelectric actuation and piezoresistive position sensing through a Wheatstone bridge. The natural frequency, measured with a vibrometer, is $f_0=93,688$ KHz.

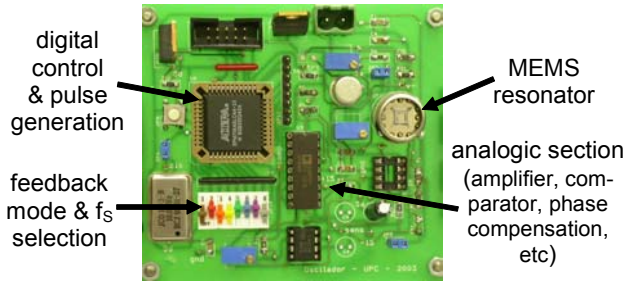

*Figure 6. Upper view of the PDO circuit used in the measurements.*

Due to the thermoelectric actuation of the resonator, force pulses are short voltage pulses provided by open-drain power transistors, while the sign detector role is played by an instrumentation amplifier plus a comparator. Digital control, sampling frequency selection, single and double feedback selection, number of delays and overall timing have been implemented on an isp CPLD. The positive values of force {+2F, +F} for the double feedback case are obtained by modulating the time length of the voltage pulses (1 or 2 µs), instead of their maximum value. In order to work with non-negligible damping values, all measurements were made in air and at room temperature.

**4.2. Results and discussion**

Figure 7.a shows an oscilloscope screen capture of the resonator position, obtained after amplification, and the set of voltage pulses applied to the MEMS excitation transistors for a single feedback topology with *m=1* (see Figure 1) and a 'perfect' sample ratio $f_0/f_S=0,25$. Let us note that, accordingly to our previous results, an excellent and stable sinusoidal waveform and a regular series of bits of the form '110011 ...' are obtained for this case.

Figure 7.b shows the results for the same case as in Figure 7.a but for a double feedback topology. We can see there that the MEMS input pulses follow the fixed pattern '12101210 ...' and that the resulting position waveform is the same one as in Figure 7.a but with a slight amplitude increase. Moreover, the bit stream spectrums corres-ponding to the single and double feedback cases look exactly the same (see Figure 8). The peak value is located at $f_{OSC}=93,68$ KHz.

In good agreement with the conclusions of the previous sections, these results tell us that no significant advantages are obtained when the double feedback is applied to the sample ratio $f_0/f_S=0,25$, which corresponds to a 'perfect' frequency case when working with the single feedback structure. We can also notice that the energy efficiency in the double feedback case is poorer: it provides more energy to the MEMS than in single feedback to obtain the same output response.

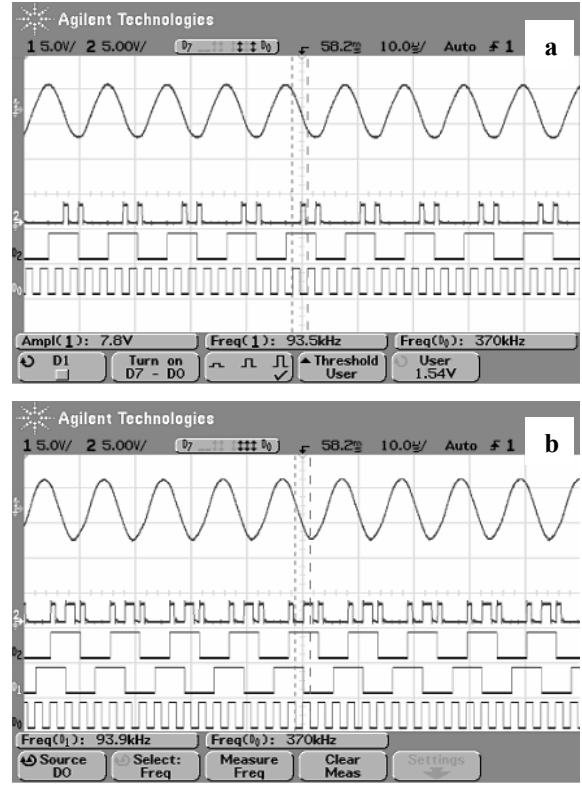

*Figure 7. Oscilloscope screen captures of resonator position (1), MEMS input pulses (2), delayed comparator output (D2,D1) and sample clock (D0), for $f_S=4f_0$ for single (a) and double (b) feedback structures.*

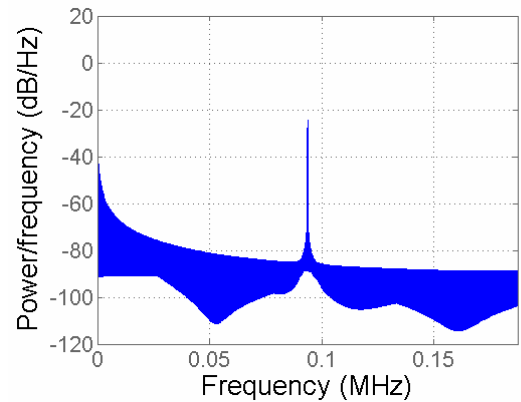

*Figure 8. Bit stream spectrum corresponding to the same experimental conditions as in Figure 7 for both single and double feedback cases.*

Again according to the conclusions of previous section, benefits of using the double feedback scheme should be more appearent for sample ratios closer to the Nyquist limit. To this effect, measurements for a sample frequency $f_S=2,489f_0$ have been carried out. This values generates an almost chaotic behaviour (no regular waveforms, no fixed bit patterns, bad spectrum, ...) when using the single feedback structure.

On the other hand, Figure 9 shows that the double feedback structure exhibits a fair senoidal waveform, plus a stable and regular '111100111100 ...' input pattern for the same $f_S=2,489f_0$ case. The corresponding spectrum, shown in Figure 10, is also very good, with a peak value located at $f_{OSC}=93,66 KHz$.

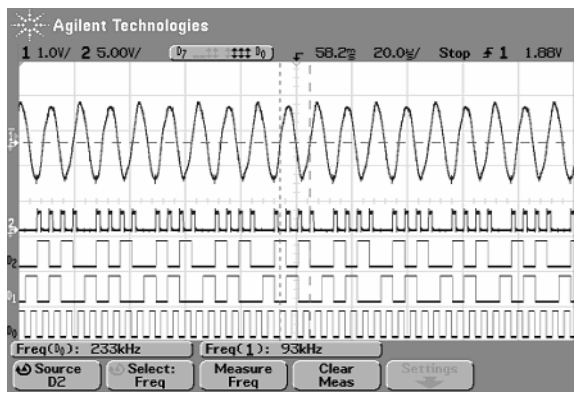

*Figure 9. Oscilloscope screen captures of resonator position (1), MEMS input pulses (2), delayed comparator output (D2,D1) and sample clock (D0), for $f_S=2,5f_0$ and a double feedback structure.*

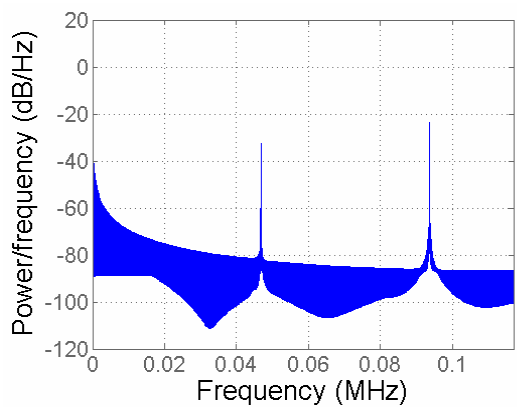

*Figure 10. Bit stream spectrum corresponding to the same experimental conditions as in Figure 9.*

## 5. CONCLUSIONS

Two different feedback loops for PDO structures have been compared. Simulation results show that the double feedback forces the oscillator to put the oscillation frequency closer to the natural frequency of the MEMS resonator, even in the case of heavy damping factors. Experimental results confirm that the double feedback architecture reduces the influence of the resonator damping losses and it also allows working in wider frequency ranges than the single feedback one, for the same number of delay blocks.